Defence Against the Modern Arts: the Curse of Statistics – FRStat

By

Cedric Neumann, Ph.D.

Associate Professor of Statistics

South Dakota State University

Brookings, SD 57007

+1 415 272 67 52 (do not publish)

Cedric.Neumann@me.com

## Abstract

For several decades, legal and scientific scholars have argued that conclusions from forensic examinations should be supported by statistical data and reported within a probabilistic framework. Multiple models have been proposed to quantify the probative value of forensic evidence. Unfortunately, several of these models rely on ad-hoc strategies that are not scientifically sound. The opacity of the technical jargon that is used to present these models and their results and the complexity of the techniques involved make it very difficult for the untrained user to separate the wheat from the chaff. This series of paper is intended to help forensic scientists and lawyers recognise issues in tools proposed to interpret the results of forensic examinations. This paper focuses on the tool proposed by the Latent Print Branch of the U.S. Defense Forensic Science Center (DFSC) and called FRStat. In this paper, I explore the compatibility of the results outputted by FRStat with the language used by the DFSC to report the conclusions of their fingerprint examinations, as well as the appropriateness of the statistical modelling underpinning the tool and the validation of its performance.

## 1. Introduction

The Latent Print Branch of the U.S. Defense Forensic Science Center (DFSC) has recently made significant changes in their operating procedures to account for numerous criticisms directed at the field of fingerprint examination over the past decades. In particular, in 2015, the DFSC issued an information paper where it announced its decision to move away from the categorical decision scheme of identification, exclusion and inconclusive examination. Instead, the Latent Print Branch proposed to use a more probabilistic language (U.S. Department of the Army, 2015):

> *"The latent print on Exhibit ## and the record finger/palm prints bearing the name XXXX have corresponding ridge details. The likelihood of observing this amount of correspondence when two impressions are made by different sources is considered extremely low."*

This language was modified in 2017 to reflect the use of two alternative propositions rather than a single likelihood (U.S. Department of the Army, 2017):

> *"The latent print on Exhibit ## and the standards bearing the name XXXX have corresponding ridge detail. The probability of observing this amount of correspondence is approximately ## times greater when impressions are made by the same source rather than by different sources."*

The DFSC has developed a tool, FRStat (Swofford et al., 2018), which attempt to provide statistical support for fingerprint examination conclusions expressed using this language. This tool has been introduced in U.S. military court and has been offered in a recent civilian case in Texas (Ex Parte Areli Escobar, 2019).

In this paper, I discuss the DFSC proposed language, FRStat's intrinsic validity and the adequacy of FRStat to support the DFSC proposed language. I focus on four main issues associated with this project, namely: the lack of compatibility between FRStat and the DFSC language, the lack of meaning of the ratio returned by FRStat, the inappropriate statistical modelling underpinning FRStat calculations, and the issues related to the validation of FRStat's performance.

In the next section, I briefly describe some technical terms commonly used in statistics and probability theory; in Section 3, I explain the algorithm behind FRStat and the correct interpretation of its outputs; and finally, in Section 4, I describe the issues associated with the DFSC language and FRStat.

## 2. Probability, likelihood and chance

In English, *probability, likelihood* and *chance* are nearly synonymous. In probability theory, these three terms are all related to uncertainty, but have specific meanings. A probability expresses the uncertainty about the realisation (in the past, present or future) of an event[1] given a fixed set of circumstances. In contrast, a likelihood is a function of the uncertainty about the circumstances that led to an observed event. In other words (and by taking some shortcuts), the use of the terms probability or likelihood is determined by whether something fixed is conditioned on something variable (likelihood) or whether something variable is conditioned on something fixed (probability). Finally, chance pertains to the parameter of a Bernoulli experiment, which means that it relates to the truth of the event (i.e., the experiment only considers whether the event is true or false; it is not concerned with multiple alternative events). (Lindley, 2007, p.115)

As an example, consider a set of observations made on a finger impression recovered from a crime scene. Assume that only one of three individuals can be the potential donor of the trace impression: Mr. X, Mr. Y or Mr. Z. In this scenario, we refer to the *probability* that Mr. X (or Mr. Y, or Mr. Z) is the donor of the trace impression given the observations made on the trace (variable donor conditioned on fixed observations). We refer to the *likelihood* of these observations given the hypothesis that Mr. X (or Mr. Y or Mr. Z) is the donor of the trace (fixed observations conditioned on multiple possible donors). Finally, we refer to the *chance* that Mr. X (vs. anybody else than Mr. X) has left the trace impression.

---

[1] An event is a set of outcomes of a situation. An event can be expressed in many forms, such as a statement (e.g., Mr. X left the trace impression) or a set of observations that can be expressed using numerical values or other types of variables (e.g., categorical variable).

Probabilities and likelihoods follow different rules:

1.  The *probability* of an event always takes a value between 0 (the event is impossible) and 1 (the event is certain);

2.  The sum of the probabilities of all possible mutually exclusive events is always 1;

3.  When the outcomes of an experiment are represented by a discrete random variable (integers or categories), the probability distribution over the possible values that the random variable can take is represented by a *probability mass function*;

4.  When the outcomes of the experiment are represented by a continuous random variable (real numbers), the probability distribution over the range of values that the random variable can take is represented by a *probability density function*;

5.  The sum of a probability mass function over all possible values of the discrete random variable is always 1. Similarly, the integral of the probability density function over all possible values of the continuous random variable is always 1;

6.  The probability of any given value for a discrete random variable is equal to the probability mass assigned to that value. However, the probability of any given value for a continuous random variable is always 0. A specific value from a continuous random variable can only be associated to a *probability density*. In the case of a continuous random variable, it is only possible to assign probabilities to a range of values[2];

---

[2] Consider an experiment that generates values between 0 and 1. The intuition is that the experiment will never generate a value exactly equal to 0.5 (or any other value). Thus, the probability that the outcome of the experiment is exactly 0.5 is 0. However, it is possible that the experiment will generate a value between 0.495 and 0.505. Thus, it is appropriate to discuss the probability that the outcome of the experiment is included in the range [0.495,0.505].

7. When the range of values considered by the probability statement goes from the lowest possible value to some value, or from some value to the largest possible value that a random variable can take, the term *tail probability* is used;

8. *Probability densities* can take values larger than 1 while *probability masses* (and *probabilities*) cannot;

9. The *likelihood* of a set of observations is equal to the joint probability mass (if discrete) or density (if continuous) assigned to the observations. Therefore, the likelihood of any value taken by a continuous random variable exists (as opposed to the probability of the value), providing that a distribution function exists for that random variable. Additionally, likelihoods can take values larger than 1 if they are functions of continuous random variables;

10. The sum of the likelihoods of an event under all possible circumstances does not, in general, sum to 1.

The distinction between probability, tail probability, likelihood and probability density will become critical when as we discuss FRStat and the associated language from the DFSC in the next sections.

### 3. FRStat

FRStat is an algorithm, encapsulated in a software package, developed by the DFSC to support the probabilistic language announced in 2015 and 2017. The overall purpose of FRStat is to calculate the risks of erroneous identification and erroneous exclusion for any given comparison between a latent and a control impression. FRStat was the subject of a publication by Swofford et al. (2018)[3].

### *3.1 FRStat overview*

FRStat is based on two premises:

1. Pairs of impressions from a given finger are more similar to each other than two impressions from two different fingers;

---

[3] All of the content from Section 3 is interpreted from Swofford et al. (2018).

2. The risk of making an erroneous identification increases when the level of similarity between a trace and a control impression decreases, while the risk of making an erroneous exclusion increases when the level of similarity increases.

Overall, FRStat is designed to estimate the risks of erroneous identification and exclusion for a given level of similarity between a trace and a control finger impression. Formally, FRStat is designed to test the two following hypotheses[4]:

$H_0$: the trace and control impressions originate from the same donor;

$H_1$: the trace and control impressions originate from two different donors.

FRStat works in a similar way to frequentist hypothesis testing. It is designed to accept or reject $H_0$ based on the risks of type I and type II errors. As such, FRStat relies on a test statistic, on the sampling distributions of that test statistic under the considered hypotheses, and on the estimation of the type I and type II error rates for a given value of the test statistic using *tail probabilities* assigned based on the sampling distributions. The two main differences between FRStat and the classic statistical hypothesis testing framework are that:

1. The nomenclature of error types is inverted in the forensic context. FRStat's type I error rate ($\alpha$) corresponds to the false negative rate since erroneously rejecting $H_0$ results in wrongly concluding that the trace and control impressions originate from two different donors. FRStat's type II error rate ($\beta$) corresponds to the rate of false positives since failing to reject $H_0$ results in erroneously concluding that the trace and control impressions originate from the same donor;

---

[4] Note that the two hypotheses are not specific to a named individual. This point is important as it influences the design of the experiment(s) aimed at addressing the hypotheses. This type of hypotheses is considered to address the *common source* identification problem in forensic science (Ommen et al. 2017, 2018).

2. Contrary to most statistical tests, FRStat estimates a rate of type II error: estimating the type II error enables accepting $H_0$ with a certain level of risk, instead of merely failing to reject it when the risk of type II error is unknown.

### 3.2 FRStat's test statistic

For a given comparison between a latent and a control impression, FRStat's test statistic is based on the level of similarity between the trace and the control impression. The first stage of FRStat's algorithm is dedicated to measuring this statistic for any pair of impressions. The manner in which FRStat measures this similarity is irrelevant for the rest of the discussion and is not studied any further. The only important point is that this stage of the algorithm returns the level of similarity between the trace and control impressions as a continuous univariate random variable (i.e., a real number). The scale and range of this number are equally irrelevant, apart from noting that larger numbers correspond to greater levels of similarity between two impressions.

### 3.3 Sampling distributions

To estimate the risks of erroneous identification and exclusion for a given level of similarity and a given number of corresponding features, FRStat relies on the sampling distributions of the similarity statistic under the two hypotheses state above, conditioned on the number of features. FRStat allows for processing fingerprint comparisons that have between 5 and 15 features. For each number of features, parametric sampling distributions of the similarity statistic are learned from two datasets, a *mated* and a *non-mated* dataset.

For each number of features, the *mated* dataset enables estimating the rate of false exclusions ($\alpha$) and is constituted of 1,996 similarity scores sampled from 499 pairs of impressions obtained from 50 individuals (10 pairs from each individual). The *mated* dataset for 15 features only includes 499 scores.

For each number of features, the *non-mated* dataset enables estimating the rate of erroneous identifications ($\beta$) and is constituted of 2,000 similarity scores sampled from approximately 100 pairs of impressions, where the impressions in each pair are from two different donors[5].

To extrapolate the rates of false positives and negatives beyond the range of observed similarity scores, FRStat uses parametric mixtures of logistic distributions to represent the *probability densities* of the similarity scores in the mated and non-mated datasets[6].

It is important to note that the distributions of scores in both datasets are fixed in the sense that they are not case specific. The same distributions based on the same scores are pre-computed and used for all casework comparisons. These distributions are not tailored to specific observations made in a given case. This implies that the risks of errors reported by FRStat are average risks, and not specific to the case at hand.

### 3.4  Type I and type II error rates

The second stage of FRStat's algorithm is concerned with estimating the type I ($\alpha$) and type II ($\beta$) error rates for the test. As in classical hypothesis testing, these error rates are *tail probabilities* assigned using the sampling distributions mentioned above. The risk of erroneous identifications (type II, $\beta$) corresponds to the probability of observing high similarity scores in the non-mated sampling distribution (right tail probability in the non-mated sampling distribution, Figure 1), while the risk of erroneous exclusions (type I, $\alpha$) corresponds to the probability of observing low similarity scores in the mated sampling distribution (left tail probability in the mated sampling distribution, Figure 1).

---

[5] It is not clear if Swofford et al. (2018) attempted to obtain non-mated scores as high as possible. Swofford et al. (2018, p.115) states that *"... the empirical distributions are intentionally biased such that the non-mated data are biased to higher similarity statistics values..."*. In addition, in the Supplemental Appendix II of Swofford et al. (2018), which is referenced in Section 2.2 of Swofford et al. (2018), we read that *"Each query print was then searched using an Automated Fingerprint Identification System (AFIS) against an operational database containing approximately 100 million different fingerprint impressions..."*. However, still in Section 2.2, Swofford et al. (2018) state that *"The distributions of similarity statistic values [...] of non-mated samples [...] were generated using a subset of impressions from the [...] NIST Special Database (SD) 27 [...] randomly paired to non-mates."*. My interpretation is that Swofford et al. (2018) consider scores from random non-mated pairs of impressions when building the distributions of the sampling statistics under $H_1$, but use the AFIS system to generate highly similar non-mated pairs of impressions when testing their model.  Naturally, proper design of FRStat would have required the authors to study the non-mated distributions of the similarity statistic using highly similar non-mated impressions to begin with.

[6] More will be said about these mixtures of logistic distributions below.

### 3.5 FRStat operational use

From my understanding of the FRStat operating procedures at the DFSC, FRStat is only used when an examiner believes that there is an *association* between a pair of trace and control impressions. FRStat is not used in cases when the examiner believes that the examination is inconclusive or would result in an exclusion. The goal of FRStat is to quantify the significance of the association.

Figure 1 represents a typical FRStat calculation. The level of similarity between the trace and the control impressions in the considered case is first calculated based on the *corresponding* features annotated by the fingerprint examiner on the images of the trace and control impressions. This level of similarity is the *observed score* and acts as the statistic that is used to test the pair of hypotheses mentioned in a previous section. FRStat then calculates the probability of observing a score *smaller* than the observed score in the hypothesis that the observed score originates from the mated score distribution (i.e, a score that would imply weaker correspondence between a trace and control impression originating from the same donor). This probability is assigned using the left tail probability of the mixture of logistic distributions fitted to the mated dataset for the considered number of features and corresponds to the risk of false exclusion, $\alpha$. FRStat also calculates the probability of observing a score *larger* than the observed score in the hypothesis that the observed score originates from the non-mated score distribution (i.e., a score that would imply stronger correspondence between a trace and control impression originating from different donors). This probability is assigned using the right tail probability of the mixture of logistic distributions fitted to the non-mated dataset and corresponds to the risk of false identification, $\beta$. FRStat finally returns a set of three numbers: $\alpha$, $\beta$ and $\frac{\alpha}{\beta}$.

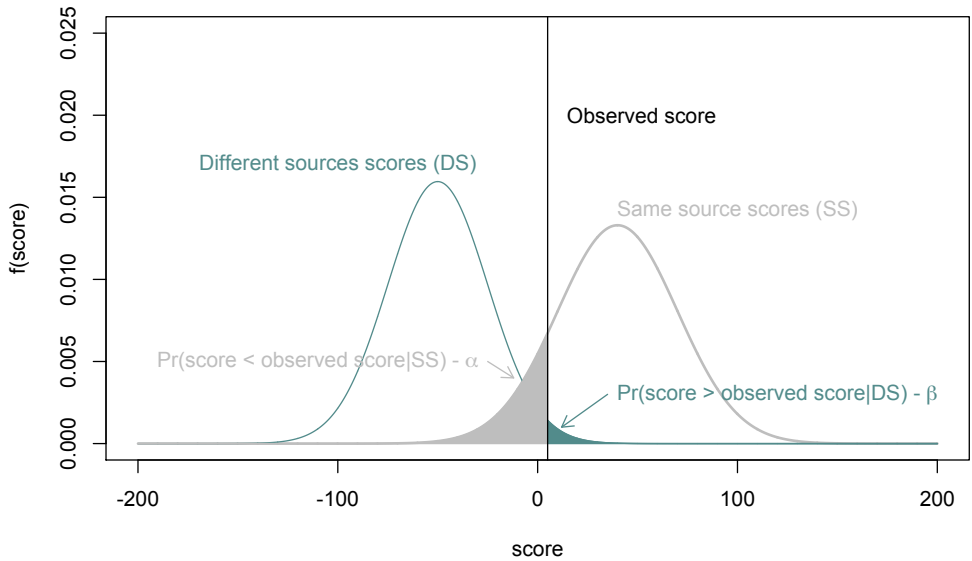

Figure 1: Representation of an FRStat calculation with indication of the two tail probabilities considered by the algorithm. In this illustration, the vertical line indicates the value of the similarity score between a pair of latent and control impression considered in a case. FRStat considers the ratio between (1) the probability of observing a score smaller than the actual similarity score in a distribution of scores obtained from pairs of impressions originating from the same source (tail probability in light grey going to the left of the observed score) and (2) the probability of observing a score greater than the actual similarity score in a distribution of scores obtained from pairs of impressions originating from different sources (tail probability in dark grey going to the right of the observed score)

### 3.6 FRStat interpretation

The interpretation of the first two numbers reported by FRStat is straightforward. The last number does not have a clear interpretation. The first number, $\alpha$, is an estimate of the average risk of erroneous exclusion that would be associated with the conclusion that the pair of impressions considered in the case at hand originates from two different individuals. In contrast, the second number, $\beta$, is an estimate of the average risk of erroneous identification that would be associated with the conclusion that the pair of impressions considered in the case at hand originates from the same person. The last number may be considered as representing the *relative risk* of false exclusion for a given risk of erroneous identification. For example, a ratio of 1,000 would indicate that it would be 1,000 times riskier to exclude than to identify the two considered impressions[7].

4. **Issues preventing the use of FRStat in casework**

*4.1  DFSC language and its compatibility with FRStat number(s)*

The DFSC language from 2017 (U.S. Department of the Army, 2017) has several issues. The first issue is that it abuses the notion of probability. As discussed in Section 2, a probability can only be assigned to a numerical random variable when that variable is discrete. However, the *amount correspondence* considered by the DFSC is a real number, at least as calculated by FRStat (Swofford et al., 2018)[8], and would always be associated with a probability of zero. Thus, at the very least, the DFSC language should consider the *probability density* of the amount of correspondence. However, since the amount of correspondence between a trace and a control impression in a given case is fixed, and this fixed value is considered under two different sets of circumstances, the DFSC language should really be using the notion of *likelihood*.

---

[7] Issues surrounding the interpretation of this last number will be discussed further later in Sections 4.1 and 4.2.

[8] Technically, it is possible to consider that the *amount of correspondence* is an *event*, in which case it would be appropriate to talk about probability. It could also be possible to consider that the *amount of correspondence* could be returned as a discrete variable (e.g., the number of minutiae in agreement), in which case, it would also be appropriate to talk about probability. However, as we will see below, these technicalities are moot since the probabilities mentioned in the DFSC language do not correspond to the probabilities calculated by FRStat regardless of which term is used.

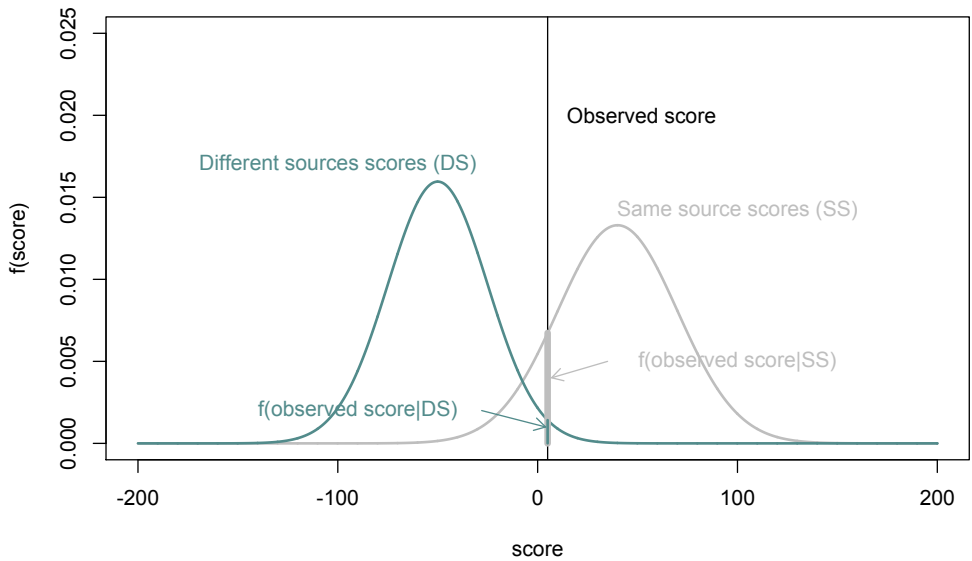

Figure 2: Representation of a *score-based likelihood* ratio, as expressed in the 2017 DFSC language, with indication of the two likelihoods considered in the ratio. In this illustration, the vertical line indicates the value of the similarity score between a pair of latent and control impression considered in a case. A score-based likelihood ratio considers the ratio between (1) the likelihood of the actual similarity score in a distribution of scores obtained from pairs of impressions originating from the same source (probability density in light grey at the value of the observed score) and (2) the likelihood of observing the actual similarity score in a distribution of scores obtained from pairs of impressions originating from different sources (probability density in dark grey at the value of the observed score).

Secondly, the DFSC language is expressing what has been called a *score-based likelihood ratio*. Score-based likelihood ratios have been proposed for fingerprint evidence by numerous authors for more than a decade (see for example Gonzalez-Rodriguez et al. 2005, Neumann et al. 2007). The likelihood ratio traditionally advocated by the legal and scientific communities is essentially a function of the similarity between the characteristics of the trace and control objects in its numerator, and a function of the rarity of the trace characteristics in its denominator. In contrast, a score-based likelihood ratio is a function of the similarity between the characteristics of the trace and control objects in both its numerator and denominator as illustrated in Figure 2. The main shortcoming of score-based likelihood ratios is that they fail to account for the rarity of the trace characteristics[9,10] and that, they are not adequate measures of support for an hypothesis over another in the forensic context.

Finally, and most importantly, despite FRStat's results being reported using the 2017 DFSC language, the numbers produced by FRStat are not compatible with this language. I explained previously that FRStat returns three numbers for any given pair of trace and control impressions (and for the associated measure of similarity between them): the risk of erroneous exclusion ($\alpha$), the risk of erroneous identification ($\beta$), and the ratio between these two risks ($\alpha/\beta$). The two risks of erroneous conclusions are *tail probabilities*: the risk of erroneous exclusion is the probability of observing a level of correspondence *smaller* than the one observed in the case at hand in the mated distribution, while the risk of erroneous identification is the probability of

---

[9] An example based on serology data illustrates the difference. Assume the distribution of ABO blood types in the U.S.A. is 44% O; 42% A; 10% B; and 4% AB. Further assume that a suspect has the same blood type as a bloodstain recovered at the crime scene. Then (1) the probability of observing "corresponding features" (the blood types) in two blood samples if they both come from the same person is 1 regardless of the blood type; (2) the probability of observing "corresponding features" in two blood samples if they come from two different individuals is $0.44^2 + 0.42^2 + 0.1^2 + 0.04^2 = 0.3816$. Note that neither of the two probability statements is conditioned on the blood type actually observed. As a result, the "weight of evidence" for the corresponding blood types is $1/0.3816 = 2.62$.

However, since the serologist observes a specific blood type, the weight of evidence should be calculated by accounting for the varying rarity of the different blood types. The probability statements of interest should really be: (1) the probability of observing a bloodstain and a suspect with blood type XXX if the bloodstain was made by the suspect; the probability of observing a bloodstain and a suspect with blood type XXX if the bloodstain was made by somebody else than the suspect. The result is then O: $1/0.44 = 2.27$; A: $1/0.42 = 2.38$; B: $1/0.10 = 10$; and AB: $1/0.04 = 25$. Therefore, the "correspondence" ratio of 2.62 overstates the weight of the evidence if the suspect's blood and the crime scene bloodstain are common (types O and A). For those types, it makes the "features" appear rarer than they are. Contrariwise, 2.62 understates the weight of the evidence if the blood types are the rare (types B and AB). It makes these "features" appear more common than they are.

[10] An extended discussion of limitations of score-based likelihood ratios will be the subject of a second paper and is summarized in Neumann et al. (2019).

observing a level of correspondence *greater* than the one observed in the case at hand in the non-mated distribution. Neither probability calculated by FRStat is the probability (or likelihood, or probability density) of observing *exactly* the level of correspondence in the case at hand as the 2017 DFSC language presents it. Graphically, the DFSC language attempts to convey the situation presented in Figure 2, which is quite different from the FRStat calculations represented in Figure 1.

An appropriate manner to convey FRStat numbers could read as follows:

> *The latent print on Exhibit ## and the standards bearing the name XXXX have corresponding ridge detail. The probability of observing a smaller amount of correspondence when impressions are made by the same source is approximately ## times greater than the probability of observing a greater amount of correspondence when impressions are made by different sources.*

While technically correct, this language retains the flavour of the likelihood ratio (which FRStat is not intended to calculate) and is convoluted to the point that it has no intuitive meaning. A more intuitive and descriptive language could read as:

> *The latent print on Exhibit ## and the standards bearing the name XXXX have corresponding ridge detail. At the observed level of correspondence between the exhibit and the standards, the risk of erroneous exclusion is ## times greater/smaller than the risk of erroneous identification.*

That said, even this language may be perceived as disingenuous since it only presents the relative risks and does not make explicit the actual risk of an erroneous identification. Hence, irrespective of the value of the ratio, it would be important for fact-finders to understand the magnitude of the risk of erroneous identification to properly interpret the ratio.

One of the key features of a likelihood ratio is that it has a "tipping point": values of the ratio above one indicate that the set of circumstances conditioning the numerator is supported by the observations, while values below one indicate that the same observations support the set of circumstances conditioning the denominator. A ratio of one represents an "inconclusive" situation where the evidence does not support any of the two proposed sets of circumstances. While the authors of Swofford et al. (2018, p118) are very clear that their ratio is not a likelihood ratio, they erroneously describe its behaviour as if it was one (Swofford et al., 2018, p116). It should be self-evident that the value of the ratio between tail probabilities is not equal to the value of the ratio between likelihoods, even when calculated at the same point. As a consequence, the two ratios may end up on two opposite sides of one, which may result in contradictory evidence. For example, Figure 3 illustrates a situation where the ratio of the two tail probabilities is exactly equal to one when the ratio of the likelihoods is not. In Figure 3, the ratio of the *likelihoods* of the similarity statistic would support the idea that the two impressions where *not* made by the same person according to the DFSC language (U.S Department of the Army, 2017), while the ratio of the two *tail probabilities* would result in inconclusive evidence according to Swofford et al. (2018).

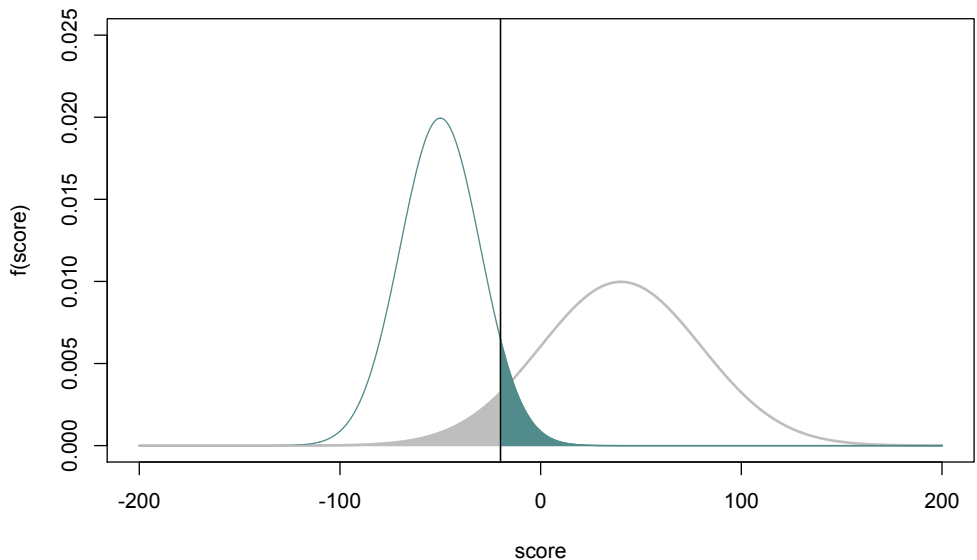

Figure 3: Representation of a situation where the ratio of the likelihoods of the observations (value given by the ratio of the y-axis values at the points where the vertical line intersects the two distributions) is well above one, while the ratio of the two tail probabilities (value represented by the ratio of the shaded areas) is exactly one.

It seems intuitive that a large ratio (i.e., high risk of erroneous exclusion over a low risk of erroneous identification) would tend to favour the notion that a pair of impressions was made by the same person. However, it is logically incorrect to imply that an FRStat ratio above one *supports* the hypothesis of common source, while a ratio below one *supports* the alternative hypothesis (Swofford et al., 2018, Equation 3). In fact, there is no rule that states that the "tipping point" of the ratio of the rates of erroneous identifications and exclusions should be one. Such a rule would imply that an erroneous exclusion is as serious as an erroneous identification. Any attempt to set the value of the "tipping point" of FRStat would require a general discussion between the scientific, legal and political communities in order to determine the appropriate relative severity between these two risks[11].

---

[11] A parallel can be made between the ratio of the rates of erroneous identifications and exclusions and the ratio of the risks of convicting an innocent and freeing a guilty. The determination of the relative severity in the latter case is an unsolvable problem.

The general convergence between the FRStat ratio and the *specific source* likelihood ratio (Ommen et al., 2017, 2018) has been studied by Neumann et al. (2019). Some results obtained using a toy example are reproduced in Figure 4.

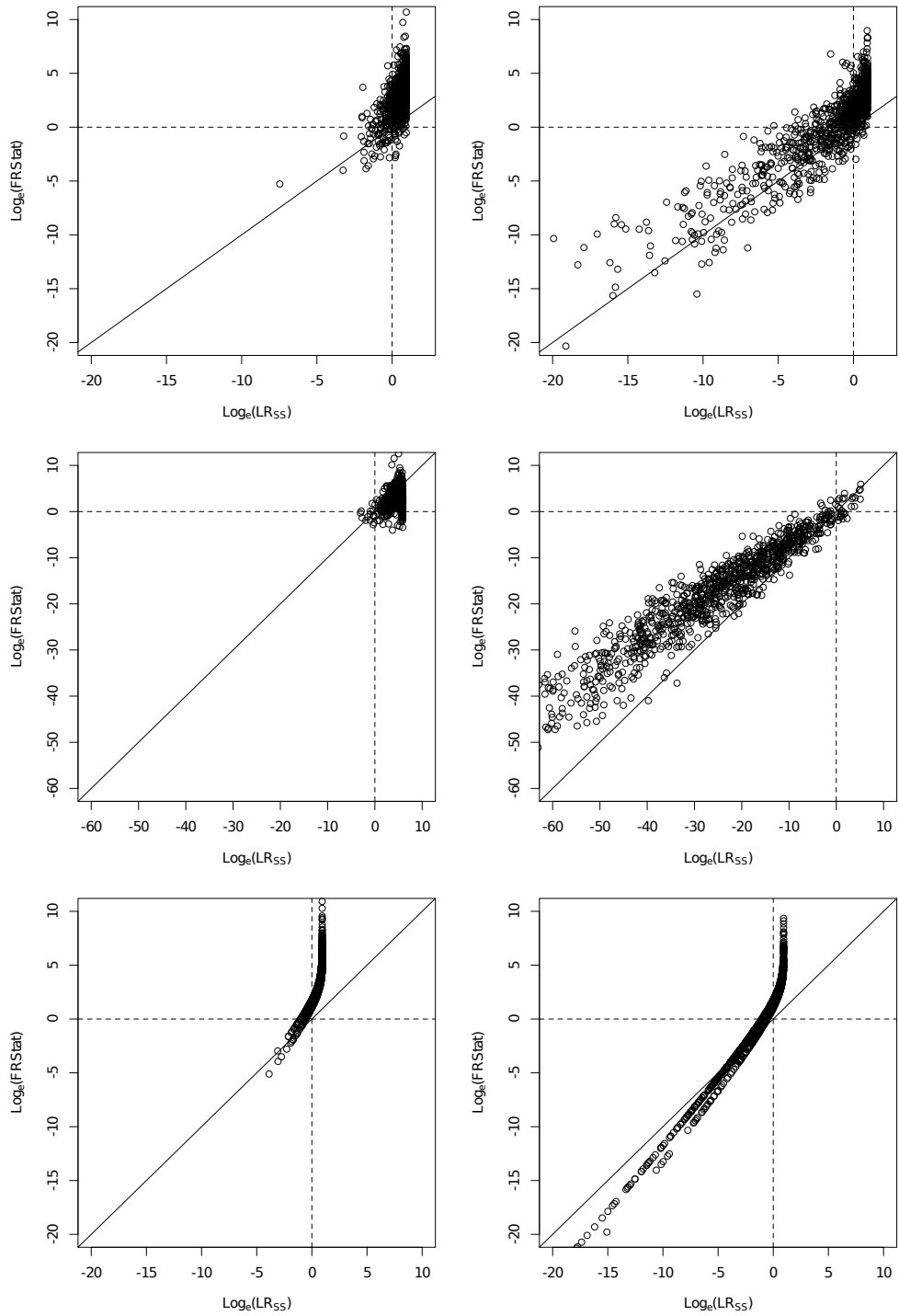

Figure 4: Comparisons between FRStat-like values with LRs in the specific source scenario. Columns: the left column reports the results when the observations are sampled when H0 is true; the right column reports the results under H1. Rows: (a) the source of the control impression is common and has some variance; (b) the source of the control impression is rare and has some variance; (c) the source of the control impression is common and has virtually no variance. The data originates from 1,000 simulations using normal distributions, with which it is trivial to calculate FRStat-like numbers and compare them to the corresponding true likelihood ratios.

The left column of Figure 4 shows the convergence of the reported FRStat number (y-axis) to the likelihood ratio (x-axis) when the sets of observations truly originate from the same source, while the right columns shows the convergence when the sets of observations truly originate from different sources. The first row of Figure 4 shows the convergence when the characteristics of the suspected source are common and variable; in the second row, the characteristics of the suspected source are rare and variable; and in the last row, the characteristics of the suspected source are common and have virtually no variability. Figure 4 shows that in no circumstance does the FRStat number and the likelihood ratio converge. In fact, as a general rule, the FRStat number overstates the weight of the evidence and, in most situations, the evidence would be highly misleading in favour of the prosecution's case. This may not be of any importance when a pair of latent and control impressions truly originate from the same donor; however, this behaviour will have dramatic consequences for innocent suspects.

### 4.3 Statistical underpinning of the FRStat algorithm[12]

The purpose of FRStat's algorithm is to assign the rates of false exclusions and false identifications by modelling the tail of the probability distributions of the scores in the mated and non-mated datasets (Section 3). Figure 5 in Swofford et al. (2018) shows that these two distributions are well separated upwards of 9 features, but that the non-mated score distribution has a consistent trailing right tail, indicating that high score values for non-mated pairs of impressions are not uncommon. The plot for 15 features in Figure 5 from Swofford et al. (2018) is reproduced below as Figure 5.

FRStat is intended to be primarily used when an examiner believes that two impressions are sufficiently similar that the common source hypothesis cannot be excluded. Thus, any given comparison processed through the FRStat algorithm is likely to return a high similarity score, regardless of the true origin of the impressions. To avoid implicating an innocent suspect and, by extension, executing miscarriages of justice, it is critical that FRStat be able to provide accurate information on how often such high scores are expected to be observed when the pairs of impressions are not from the same source. For example, Figure 5 shows that there is a non-trivial amount of non-mated high scores with five out of 2,000 (0.25%) scores being larger than zero for 15 features. The point is that the modelling of the mated score distribution (henceforth the rate of false exclusion) is not that critical overall, as opposed to the modelling of the non-mated score distribution, and in particular of its right tail which is used to assign the rate of erroneous identification.

---

[12] All data and information supporting the statistical analysis in this section were obtained as part of the discovery material in the Ex Parte Areli Escobar case (Ex Parte Areli Escobar, 2019). The data are available to the author of this paper but should be requested directly to the authors of Swofford et al. (2018).

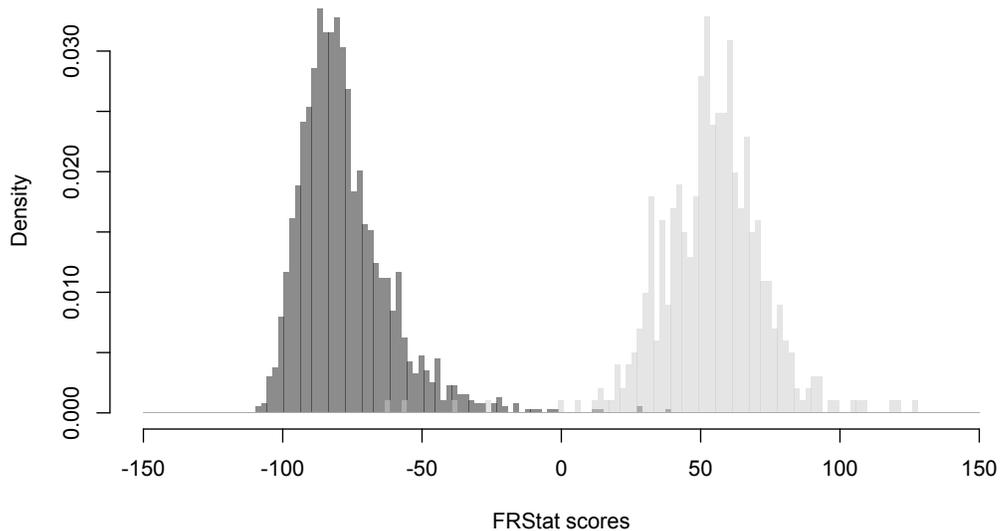

Figure 5: Reproduction of Figure 5 from Swofford et al. (2018). Minor differences between these figures are due to the use of different bin sizes for the histograms. Same source scores ($N$=499) are shown in light grey (right distribution). Different sources scores ($N$=2,000) are shown in darker grey (left distribution). The figure shows two trailing tails of low similarity scores between same source impressions and high similarity scores between different sources impressions.

The authors of Swofford et al. (2018) have chosen to model their score distributions using mixtures of logistic distributions. Logistic distributions are unimodal and symmetrical distributions that have fatter tails than Gaussian distributions and will give higher probability density to rarer observations (hence making them appear less rare). The general idea behind the choice of using logistic distributions seems to be that fatter tails will provide larger estimates for the risk of erroneous identification and, therefore, will consistently result in smaller FRStat ratios. This strategy can be perceived as more favourable to an accused. The key question is whether this strategy is sufficient to represent the true rate at which larger similarity scores can be observed between pairs of impressions that are not from the same source.

Figure 6 shows the model chosen by Swofford et al. (2018) to represent the distribution of scores for the non-mated distribution of scores for 15 features used in FRStat and represented in Figure 5, together with a histogram of the data used to estimate the parameters of the model (the data is the same as in Figure 5 above). The left panel of Figure 6 superimposes the model (solid line) and the data (histogram) and it appears that the model is reasonably appropriate for the body of the data. The right panel of Figure 6 shows the same data as in the left panel using a logarithm transform on the y-axis. This logarithm transform allows to better observe data points that have a rarer occurrence. The right panel shows that while the probability density function "dives" in the right rail, the relative frequency of the actual observations remains steady, even for fairly high similarity scores between non-mated pairs of impressions. This discrepancy between the model and the data ultimately results in tail probabilities that significantly underestimate the risk of erroneous identification in casework.

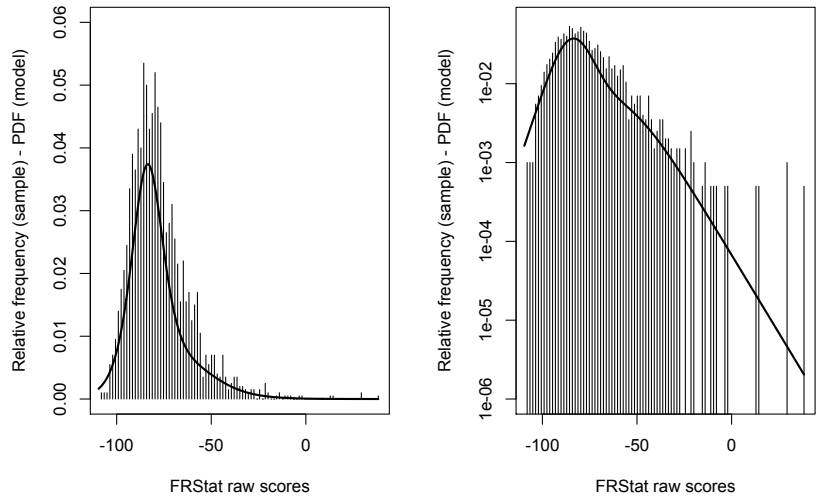

Comparison between sample and reference model
Non-mated - 15 features

Figure 6: Comparison between the observed distribution of non-mated scores for 15 features (histogram) and the mixture of logistic distributions chosen to represent the data (solid line). The parameters of the mixture of two logistic distributions for the non-mated scores for 15 features used by Swofford et al. (2018) are: weights: 0.8, 0.2; locations: -83.75, -61.25; and scale: 5.625, 10.9375. The relative frequency of the scores is shown in absolute scale (left panel) and logarithmic scale (right panel).

In addition to Figure 6, it is possible to compare the expected number of scores based on the mixture of logistic distributions used by FRStat and the actual number of scores observed by Swofford et al. (2018) during their experiments. Table 1 compares these expected and observed numbers of scores for the non-mated distribution of scores for 15 features. The observed number of scores for 15 features include all scores calculated by Swofford et al. (2018) to support Figure 5 and Tables 4, 5a and 5b of their paper. This represents all non-mated scores obtained for 15 features by the research team and reported in their paper[13].

---

[13] Figure 5 in Swofford et al. 2018 has 5 out of 2,000 scores above 0; Table 4 has 0 out of 500 scores above 0; Table 5a has 6 out of 100 scores above 0; and Table 5b has 24 out of 100 scores above 0. Some scores up to 54 have been observed. Combining all these scores collected under different conditions (see Footnote 5 in Section 3.3) may be questionable. However, (1) this is the best estimate of the relative frequency of high similarity scores for non-mated pairs of impressions that one can have based on Swofford et al. (2018), and (2) the present exercise is merely intended to indicate that, in the typical casework situation where

Table 1 shows that the mixture of logistic distributions significantly underestimates the number of high similarity scores between pairs of impressions from different donors. This means that FRStat fails to appropriately account for the probability of observing high similarity scores between a latent print and the control impression from an innocent suspect. This is particularly worrying given that FRStat is only meant to be used when an examiner cannot exclude a person as the source of a latent impression, which will only happen in case of highly similar prints. In these cases, the current modelling of the tails of the non-mated score distributions by FRStat will be extremely prejudicial to potentially innocent suspects.

| DS scores greater than | 0 | 25 | 50 |
|---|---|---|---|
| FRStat expected number of scores based on chosen model | 73 in 100,000 (0.073%) | 7 in 100,000 (0.007%) | 0.7 in 100,000 (0.0007%) |
| Observed number of scores | 1,300 in 100,000 (35 in 2,694 or 1.29%) | 519 in 100,000 (14 in 2,694 or 0.51%) | 111 in 100,000 (3 in 2,694 or 0.11%) |

Table 1: Comparison between expected number of scores between pairs of impressions originating from different donors (with 15 features in common) and the relative frequency of scores observed by Swofford et al. in their 2018 paper. The numbers in parentheses are the numbers of scores actually observed. They have been rescaled to a common scale for easier comparison between expected and observed scores.

There are three main issues that explain the lack of suitability of the FRStat mixture of logistic distributions to represent the tails of the score data. The first issue is that Swofford et al. collected non-mated scores by randomly associating latent and control impressions (Swofford et al., 2018, p155, see Footnote 5 above) instead of selecting pairs of highly-similar non-mated impressions. The second issue is that, despite having fatter tails than a Gaussian distribution, the tails of the logistic distribution are still governed by the natural exponential function, which rapidly drives the densities in the tail to zero for values of the random variable that are away from the mean. Thus, this distribution is not adequate to model the long flat tails that are

---

FRStat is intended to be used, FRStat tail modelling will vastly underestimate the expected occurrence of large similarity scores between non-mated pairs of impressions (in other words, this exercise is not meant to quantify the lack of fit).

observed in the data (see Figures 5 and 6). The last issue is that Swofford et al. (2018) rely on an inappropriate test to assess the quality of the fit of the mixtures of logistic distributions. Swofford et al. (2018) learn the parameters of the logistic distributions for the mated and non-mated datasets for each number of features (from 5 to 15) by maximum likelihood estimation using 75% of the score data. The quality of the fit between the observed similarity scores and the parametric model is tested on the remaining 25% of the score data using the Kolmogorov-Smirnov distance between the empirical and parametric distributions. The equation of the Kolmogorov-Smirnov test statistics for a proposed cumulative distribution function $F(x)$ and an empirical distribution $F_n(x)$ of $n$ observations is reproduced in Equation 1:

$$D_n = \sup_x |F_n(x) - F(x)| \tag{1}$$

It is clear from Equation 1 that the test statistic is driven by the maximum absolute distance between the proposed and observed distributions. This maximum absolute distance will always live in the body of the distribution and never in the tails. This explains the insensitivity of the Kolmogorov-Smirnov test to the lack of fit in the tails of the distributions. Swofford et al. (2018) could have used a different distance, such as the Anderson-Darling distance:

$$A_n^2 = \int_{-\infty}^{\infty} \frac{(F_n(x) - F(x))^2}{F(x)(1 - F(x))} dF(x) \tag{2}$$

As Equation 2 shows, the Anderson-Darling statistic is a weighted average of all differences between the proposed and observed distributions. Mechanically, the Anderson-Darling statistic gives heavier weights to differences between the tails of the observed and proposed distributions[14]. The lack of power of the Kolmogorov-Smirnov test can be seen in Figure 7. It is a well-known fact that the distribution of the p-value for a statistical test, when the null hypothesis is true, is uniform. Figure 7 compares empirical distributions of the p-values for four different tests obtained by simulation using the score data supporting FRStat algorithm. Each simulation ran as follows:

1. A dataset containing 75% of the non-mated score data from FRStat was created by randomly sampling from the same 2,000 observed scores as Swofford et al. (2018);

2. The parameters of the mixture of logistic distributions for the score distribution were learned by maximum likelihood estimate;

3. A Kolmogorov-Smirnov test was performed using the proposed mixture of distributions and the remaining 25% of the observed scores. The p-value was stored;

4. An Anderson-Darling test was performed using the proposed mixture of distributions and the remaining 25% of the observed scores. The p-value was stored;

5. A dataset of 1,500 scores was resampled from the proposed mixture of distributions obtained in 2;

6. A Kolmogorov-Smirnov test was performed using the proposed mixture of distributions and the scores resampled from that mixture. The p-value was stored;

7. An Anderson-Darling test was performed using the proposed mixture of distributions and the scores resampled from that mixture. The p-value was stored;

8. Steps 1. through 7. were repeated 1,000 times.

---

[14] When $F(x) = 0.5$, $F(x)(1 - F(x)) = 0.25$ and the weight is $\frac{1}{F(x)(1-F(x))} = 4$. Any other value of $F(x)$ will result in a larger weight. For example, when $F(x) = 0.99$ (i.e., probability in the right tail is 0.01), $F(x)(1 - F(x)) = 0.0099$ and the weight is $\frac{1}{F(x)(1-F(x))} = 101.01$

Steps 1 through 3 reproduce exactly what Swofford et al. (2018) did. Step 4 performs the same test as they did, but with the more powerful Anderson-Darling statistic. Steps 5 through 7 aim at studying the distributions of the p-values when the null hypothesis that the observed data truly originate from the proposed distribution is true with both tests. Since the dataset supporting FRStat is finite, repeating the simulation from steps 1 through 7 does not result in independent p-values. Therefore, some departure from uniformity is expected. The upper left panel in Figure 7 shows the distribution of the p-values for step 3. This distribution represents the distribution of the p-values from the experiment in Swofford et al. (2018). This panel shows some departure of uniformity (which is expected given the lack of independence of the simulations) and shows that Swofford et al. (2018) had a fair chance of observing a p-value greater than 0.05 when performing their test of adequacy.

That said, the upper left panel can be compared with the upper right panel, which shows the results from the more powerful test in step 4. of the simulations, and the lower left panel, which presents the results from step 6. Comparing these panels indicates that the lack of uniformity of the p-value distribution in the upper left panel may not be the sole product of the lack of independency of the simulations (since the distribution in the lower left panel is much more uniform) but may result from the proposed and observed distributions being incompatible. It also highlights the lack of power of the Kolmogorov-Smirnov test to detect the departures of the proposed distribution in the tails of the observed score distribution (since the null hypothesis of equal distributions is clearly wrong based on the Anderson-Darling test).

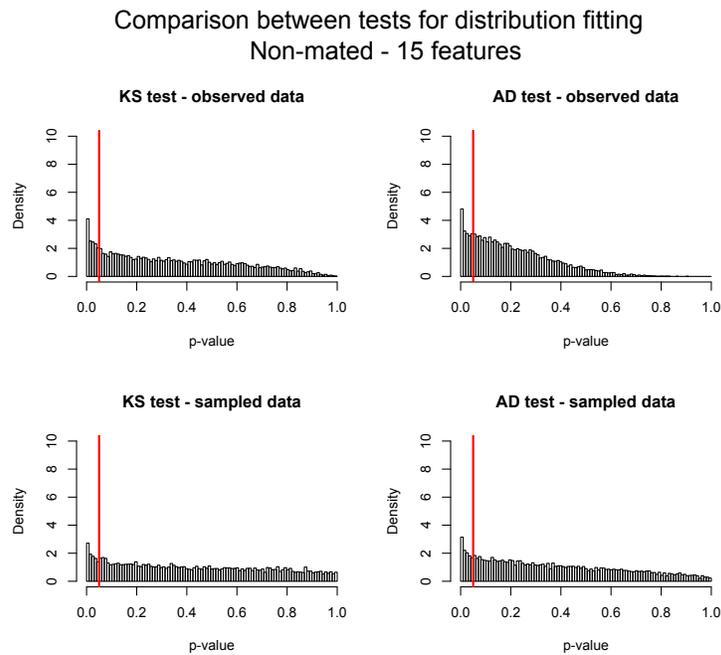

Figure 7: Results from the simulations comparing the distributions of the p-values obtained using the Kolmogorov-Smirnov (KS) test and the Anderson-Darling (AD) test. The vertical line indicates the significance level of 0.05 used by Swofford et al. (2018).

### 4.4 Validation of the performances of the FRStat algorithm.

The previous sections report the main issues associated with the calculation, interpretation and reporting of the FRStat number. From the arguments in these sections, it is clear that the FRStat number does not have the interpretation or behaviour of a likelihood ratio, that it cannot be reported using the DFSC language, and that its calculation is plagued by a lack of adequacy between the chosen parametric distributions and the observed data.

Nevertheless, it is possible to consider using the FRStat number as a purely deterministic rule. For example, it is possible to consider reporting all comparisons that produce a number above 100 as identifications[15], and all others as inconclusive (with the understanding the use of FRStat is limited to comparisons when the commonality of the origin of the pair of impressions cannot be excluded). Using this rule, it is possible to estimate the rate of erroneous identifications (i.e., the number of pairs of impressions that originate from different donors and show an FRStat number greater than 100) and the rate of missed identifications (i.e., the number of pairs of impressions that originate from the same donor and show an FRStat number smaller than 100). In this context, the calculation and actual value (or magnitude) of the FRStat number are irrelevant to the rule. The only important pieces of information that need to be reported are the calculated number in the case at hand, the value of the decision threshold that defines the rule, and the rates of erroneous identifications and missed identifications for that decision threshold.

Tables 1, 4, 5a and 5b in Swofford et al. (2018) propose such data[16]. Table 1 in Swofford et al. (2018) shows estimates for the rate of missed identifications for different values of the decision rule (i.e., >1, >10 and >100). Tables 4, 5a and 5b in Swofford et al. (2018) report the estimates for the rate of correct exclusions for these values of the decision rule (i.e., <1, <10 and <100). The estimates in Table 4 are calculated using a dataset of 20 latent prints and 25 non-mated control impressions, resulting in 500 dependent cross-comparisons. In the experiment resulting in Table 4, the 25 non-mated control impressions were not selected for their similarity with the 20 latent prints. The estimates in Tables 5a and 5b are calculated using between 94 and 100 latent prints (number of latent prints varies for different numbers of features). In the experiment resulting in Tables 5a and 5b, each latent impression was searched against a large fingerprint database of "*more than 100 million*

---

[15] 100 is only used as an example and should not be understood as the number that should be used in casework.

[16] Tables 4, 5a and 5b in Swofford et al. (2018) were already briefly discussed in relation to Section 4.3 and Footnote 13 above.

*different fingerprints*" (Swofford et al. 2018) and associated with the most similar non-matching candidate returned by the matching algorithm of the database.

The estimates for the rate of missed identifications in Table 1 are interesting but not all that valuable to assess the performance of FRStat. Indeed, missed identifications would be classified as inconclusive by the decision rule and most likely not reported. The estimates for the rate of erroneous identifications in Table 4 are also interesting in the sense that they are surprisingly high given the small sample size. For example, 99.4% of the 500 cross-comparisons displaying 14 features in agreement between the 20 latent prints and 25 control impressions that originate from different donors result in an FRStat number smaller than 100. This means that 3 of these 500 comparisons result in a number above 100 and would lead to erroneous identifications using the decision rule. Bearing in mind that the 20 latent impressions and the 25 control impressions were not selected for being particularly similar to each other, this ratio of 3 in 500 shows how easy it is to find constellations of 14 features that are similar enough (according to the FRStat algorithm) to result in very large FRStat numbers.

The series of experiments resulting in Tables 5a and 5b in Swofford et al. (2018) is very valuable as it tests FRStat in a situation that is intended to be similar to the casework situation where an examiner fails to exclude an innocent donor as the source of a latent print. The difference between Tables 5a and 5b relates to the area of friction ridge skin used for the experiment: Table 5a focuses on the delta region of the fingerprint pattern, while Table 5b focuses on the core region of the pattern. The data supporting Table 5a in Swofford et al. (2018) is reproduced below in Tables 2 and 3. In these tables, the results presented in Table 5a are supplemented in two ways:

1. Tables 4, 5a and 5b in Swofford et al. (2018) report the rates of correct exclusions for three different values of the decision rule: 1, 10 and 100. Table 2 below reports the rates of correct exclusions for three additional values of the rule: 1,000, 10,000 and 100,000;

2.  Tables 4, 5a and 5b in Swofford et al. (2018) only present the rates of correct exclusions for the
    different values of the decision rule. Table 3 below presents the corresponding rates of erroneous
    identifications.

| Feature quantity | Number of pairs | <1 | <10 | <100 | <1,000 | <10,000 | <100,000 |
|---|---|---|---|---|---|---|---|
| 5 | 99 | 0.566 | 0.788 | 0.980 | 1.000 | 1.000 | 1.000 |
| 6 | 99 | 0.687 | 0.747 | 0.980 | 1.000 | 1.000 | 1.000 |
| 7 | 96 | 0.688 | 0.719 | 0.896 | 1.000 | 1.000 | 1.000 |
| 8 | 99 | 0.747 | 0.788 | 0.812 | 1.000 | 1.000 | 1.000 |
| 9 | 99 | 0.818 | 0.818 | 0.828 | 0.960 | 1.000 | 1.000 |
| 10 | 97 | 0.814 | 0.835 | 0.845 | 0.897 | 1.000 | 1.000 |
| 11 | 96 | 0.802 | 0.823 | 0.823 | 0.854 | 0.958 | 1.000 |
| 12 | 98 | 0.857 | 0.867 | 0.888 | 0.898 | 0.939 | 0.990 |
| 13 | 99 | 0.899 | 0.929 | 0.939 | 0.939 | 0.960 | 1.000 |
| 14 | 100 | 0.980 | 0.99 | 0.990 | 0.990 | 0.990 | 1.000 |
| 15 | 100 | 0.920 | 0.920 | 0.940 | 0.940 | 0.940 | 0.990 |

Table 2: Reproduction of the data supporting Table 5a in Swofford et al. (2018). The results presented in Table 5a are extended with estimated rates of correct exclusions for three additional thresholds: 1,000, 10,000 and 100,000.

| Feature quantity | Number of pairs | >1 | >10 | >100 | >1,000 | >10,000 | >100,000 |
|---|---|---|---|---|---|---|---|
| 5 | 99 | 43.4 | 21.2 | 2.0 | 0.0 | 0.0 | 0.0 |
| 6 | 99 | 31.3 | 25.3 | 2.0 | 0.0 | 0.0 | 0.0 |
| 7 | 96 | 31.2 | 28.1 | 10.4 | 0.0 | 0.0 | 0.0 |
| 8 | 99 | 25.3 | 21.2 | 18.8 | 0.0 | 0.0 | 0.0 |
| 9 | 99 | 18.2 | 18.2 | 17.2 | 4.0 | 0.0 | 0.0 |
| 10 | 97 | 18.6 | 16.5 | 15.5 | 10.3 | 0.0 | 0.0 |
| 11 | 96 | 19.8 | 17.7 | 17.7 | 14.6 | 4.2 | 0.0 |
| 12 | 98 | 14.3 | 13.3 | 11.2 | 10.2 | 6.1 | 1.0 |
| 13 | 99 | 10.1 | 7.1 | 6.1 | 6.1 | 4.0 | 0.0 |
| 14 | 100 | 2.0 | 1.0 | 1.0 | 1.0 | 1.0 | 0.0 |
| 15 | 100 | 8.0 | 8.0 | 6.0 | 6.0 | 6.0 | 1.0 |

Table 3: Estimated rates of erroneous identifications (in %) corresponding to the rates of correct exclusions presented in Table 5a in Swofford and al. (2018) and in Table 2 above.

When latent and control impressions from different donors are selected for their high degree of similarity,

Tables 2 and 3 show that the rates of erroneous identifications are far superior to the estimated average rate of erroneous identifications of the current fingerprint examination process, which is around 1% (Ulery et al., 2011, Ausdemore et al., 2019). In addition, Tables 2 and 3 indicate that these rates remain high until extremely high values are used for the decision threshold. For example, the estimates for the rate of erroneous identification for configurations of 15 features is 6% at a value of the FRStat ratio of 10,000 and 1% at a value of the FRStat ratio of 100,000. The data in Tables 2 and 3, when combined with the rate of missed identifications and FRStat number ranges in Swofford et al. (2018)'s Table 1 and Figure 7, suggest that low false positive rates can only be achieved for values of the decision threshold that are so high that the rates of missed identifications become very close to 100%. Overall, these observations reveal that FRStat would be a much more inefficient and risky deterministic decision rule than the opinion-based fingerprint examination process that it is meant to complement. This would naturally question its overall usefulness.

### 5. Conclusion

For several decades, legal and scientific scholars have argued that conclusions from forensic examinations should be supported by statistical data and reported within a probabilistic framework. Multiple models have been proposed to quantify the probative value of forensic evidence. Unfortunately, several of these models rely on ad-hoc strategies that are not scientifically sound. The opacity of the technical jargon that is used to present these models and their results and the complexity of the techniques involved make it very difficult for the untrained user to separate the wheat from the chaff. This series of paper is intended to help forensic scientists and lawyers recognise issues in tools proposed to interpret the results of forensic examinations. This paper focuses on the tool proposed by the Latent Print Branch of the U.S. Defense Forensic Science Center (DFSC) and called FRStat.

In this paper, I explore the compatibility of the results outputted by FRStat with the language used by the DFSC to report the conclusions of their fingerprint examinations, as well as issues with the statistical modelling and validation of the tool itself. My findings are as follows:

1. The DFSC language does not use appropriate probabilistic language. In fact, the 2017 DFSC language abuses several notions of probability theory and is technically meaningless. In addition, the DFSC language does not reflect the numbers outputted by FRStat and cannot be used to report them. In this paper, I suggest using following more appropriate language:

   *"The latent print on Exhibit ## and the standards bearing the name XXXX have corresponding ridge detail. At the observed level of correspondence between the exhibit and the standards, the risk of erroneous exclusion is ## times greater/smaller than the risk of erroneous identification."*

2. While Swofford et al. (2018) are very clear that their algorithm is not intended to calculate likelihood ratios, they claim that the FRStat ratio has the same properties. In this paper, I explain that this is not the case and that the FRStat ratio neither converges to a likelihood ratio nor has the same logical properties.

3. The probability distributions underpinning the calculations made by the FRStat algorithms are not modelled appropriately. The mixtures of logistic distributions used by Swofford et al. (2018) may put more probability mass on larger/smaller values of their similarity score than the Gaussian distribution would, but this is not enough. In particular, I found that these mixtures of logistic distributions significantly underrepresent the likelihood of observing large similarity score values between pairs of impressions made by different donors, and thus, are severely underestimating the risk of erroneous identification reported by FRStat for any given comparison.

4. The study of the performance of FRStat conducted by Swofford et al. (2018) shows that the overall performance of FRStat is worse than the subjective comparison and decision-making method that it is meant to supplement. In particular, FRStat is subject to high rates of erroneous identifications, even for comparisons bearing many features in agreement (according to the FRStat algorithm).

FRStat is an interesting and innovative tool. Swofford et al. (2018) propose a strategy that diverge from the commonly advocated Bayesian approach. Furthermore, FRStat is encapsulated in a software package that can readily be used by fingerprint examiners. It would be interesting to widely distribute FRStat in the community to enable discussions around standard operating procedures involving fingerprint statistical models. These discussions could focus on whether statistical models have to be used before or after an opinion has been reached by an examiner; on how to handle different results arising from the use of the model on the same comparison by two examiners; or on how to report statistical results to fact-finders? In addition, FRStat could be used to study the benefits and limitations of various workflows. For example, should a statistical model be used on every comparisons, or only on borderline ones; or more generally, how can statistical models be used to gain efficiency?

That said, my findings show that it is dangerous to use FRStat, as it stands, in casework as it will rapidly lead to miscarriages of justice.